\documentstyle[12pt,epsbox,TR97]{article}
\input{standard92.mac}
\input{local.mac}
\input{symbol.mac}

\begin{document}

\mytitle{0007}{%
Hard Instance Generation for SAT}{%
Satoshi Horie and Osamu Watanabe}{%
Department of Computer Science,
Tokyo Institute of Technology\\
({\tt watanabe@titech.ac.jp})}

\myabst{%
We consider the problem of generating hard instances
for the Satisfying Assignment Search Problem (in short, SAT).
It is not known whether SAT is difficult on average,
while it has been believed that
the Factorization Problem (in short, FACT) is hard on average.
Thus,
one can expect to generate hard-on-average instances
by using a reduction from FACT to SAT.
Although the asymptotically best reduction is obtained
by using the Fast Fourier Transform \cite{schonhage-strassen:Comput71}
(in short, FFT),
its constant factor is too big in practice.
Here we propose to use the Chinese Remainder Theorem
for constructing efficient yet simple reductions from FACT to SAT.
First by using the Chinese Remainder Theorem recursively,
we define a reduction that produces,
from $n$ bit FACT instances,
SAT instances in the conjunctive normal form with $O(n^{1+\epsilon})$ variables,
where $\epsilon>0$ is any fixed constant.
({\it Cf.}
The reduction using FFT yields
instances with $O(n\log n\log\log n)$ variables.)
Next we demonstrate the efficiency of our approach
with some concrete examples;
we define a reduction
that produces relatively small SAT instances.
For example,
it is possible to construct SAT instances
with about 5,600 variables
that is as hard as
factorizing 100 bit integers.
({\it Cf.}
The straightforward reduction yields SAT instances with 7,600 variables.)}


\smallsection
Introduction

The satisfiability problem (SAT)
is a central problem
in various fields of computer science.
Precisely speaking,
we consider the following ``search problem'':
For a given propositional Boolean formula,
find an assignment of values to the propositional variables
so that the formula evaluates to true.
This paper investigates
the way of generating hard SAT instances.
(In this paper,
we consider only ``positive'' instances,
namely,
satisfiable Boolean formulas.
Also we consider only conjunctive formulas;
a formula may be a {\em $k$-conjunctive normal form formula},
i.e.,
a conjunction of disjunctions of $k$ (or less) literals,
or it may be an {\em $k$-extended conjunctive form formula},
i.e.,
a conjunction of finite functions on $k$ (or less) variables.)

While it has been known that SAT is NP-hard,
we do not know\footnote{
There have been quite a lot investigations for solving SAT,
and we have made important observations on the hardness of SAT
(see, e.g., \cite{dimacs96})
Nevertheless,
our knowledge is far from satisfiable one.}
so much about its concrete hardness.
This contrasts to
the factorization problem (FACT),
i.e.,
the problem of computing the prime factorization of a given number.
While we do not know whether FACT is NP-hard,
we have developed some knowledge on its concrete hardness
through the development of algorithms
and various experimental attacks to the problem
(see, e.g., \cite{lenstra-lenstra:handbook90}).
Here we propose an approach
for measuring concrete hardness of SAT
that uses an efficient reduction from FACT to SAT.
Theoretically,
it is clear
that FACT is polynomial-time reducible to SAT,
and that
a SAT instance $F$ generated from a FACT instance $x$
is as hard as factorizing $x$.
The goal of this paper
is to design efficient reductions
so that we can generate SAT instances
with smaller size and higher hardness.

There are two somewhat different motivations
for designing efficient reductions.

First,
with such efficient reductions,
we can generate hard SAT instances
that could be used to test
the performance of various heuristics for SAT.
In general,
it is not so easy to generate {\it good} test instances.
On the other hand,
it is easy to generate hard instances for FACT;
just generate two large prime numbers and multiply them.
Thus,
with an efficient reduction from FACT to SAT,
we can generate hard SAT instances easily.
Also,
from FACT instances,
it is easy to generate SAT instances with a unique solution;
thus,
by negating the unique solution,
we can easily generate ``negative'' SAT instances.
(In general,
``negative'' instance generation is difficult
\cite{asahiro-iwama-miyano:dimacs96}.)

Secondly,
with efficient reductions,
we can analyze the concrete hardness of SAT.
For example,
it has been widely believed that
factorizing the product of two 256 bit prime numbers is intractable.
(In fact,
even the degree of intractability has been discussed;
see, e.g., \cite{schneier94}.)
Thus,
by reducing such hard FACT instances,
we can estimate the concrete hardness of SAT.

Because of these motivations,
reductions we define must be efficient
on a certain interval of size that we are interested in.
Thus,
a simple method is more appropriate
than efficient but complicated methods.
For example,
by using the Fast Fourier Transform
(\cite{schonhage-strassen:Comput71};
see also \cite{knuth:book81}),
one can define a reduction
that yields formulas with $O(\ell\log\ell\log\log\ell)$ variables
from products of two $\ell$ bit prime numbers,
which is asymptotically the best (so far).
Unfortunately,
however,
this reduction is almost useless for our purpose
due to its large constant factor.

In this paper,
we propose one method of defining reductions,
which is based on the Chinese Remainder Theorem.
Though simple,
we show that this method gives us efficient reductions.
First,
we define a reduction
that uses the Chinese Remainder Theorem recursively
and yields formulas with $O(\ell^{1+\varepsilon})$
from products of two $\ell$ bit prime numbers,
where $\varepsilon$ is any small constant.
Clearly,
this is not the best compared with the one defined by using FFT.
But because of its small constant factor,
we may be able to use this reduction
(or, the idea of the reduction)
for generating relatively large instances,
say, formulas with 100,000 variables.
Next,
we define a reduction that works for the case $\ell\le500$.
For example,
with this reduction,
we can construct SAT instances with 5,600 variables
that are as hard as factorizing products of two 50 bit prime numbers,
which can be used as test instances \cite{dimacs96}.
({\em Cf.}
A naive reduction yields instances with 7,600 variables.)
The same reduction also yields SAT instances with 63,000 variables
that are as hard as factorizing products of two 256 bit prime numbers.
Thus,
we can conclude that
SAT instances with 63,000 variables contain
some (in fact, many) intractable instances.
({\em Cf.}
A naive reduction yields instances with 197,000 variables.)

\paragraphskip\noindent
\underline{Notations}

\noindent
Throughout this paper,
we consider,
for FACT instances,
a product of two prime numbers of the same length,
and we use $\ell$ to denote their length (i.e., the number of bits).
For any $a_{l-1},...,a_0\in\{0,1\}$,
we regard $(a_{l-1},...,a_0)$ as a binary representation of some number.
In general,
for any $a_{l-1},...,a_0\in[0,...,b-1]$,
$(a_{l-1},...,a_0)$ is a base $b$ representation of some number.

'

\smallsection
Basic Idea and Asymptotic Analysis

Here we first explain the basic idea of our method,
and then discuss the way to apply it recursively
to get an asymptotically better reduction.

Our goal is to generate,
for a given integer $x=p\times q$,
where $p$ and $q$ are $\ell$ bit prime numbers,
a SAT instance $F_x$ such that
one can easily compute $p$ and $q$
from the satisfying assignment of $F$.
In the following,
let us fix this $x$ and thus, $p$ and $q$.
Note that
$F_x$ is defined for each $x$,
and $x$ can be embedded in the definition of $F_x$ as a constant.
On the other hand,
our construction must be independent from $p$ and $q$;
in other words,
$F_x$ must be constructed without knowing $p$ or $q$.
(Otherwise,
one may extract information on $p$ or $q$ from $F_x$
without solving $F_x$.)

For our goal,
consider, for example,
$\F{1}$ that satisfies the following:

\[
[~(\veca)\times(\vecb)=x~]
~\iff~
[~\F{1}(\veca,\vecb)=\mbox{true}~].
\]

\noindent
Here $a_i$ and $b_i$ are propositional variables,
and we use them to represent nonnegative integers.
The satisfying assignment of this $\F{1}$
is the binary representation of $p$ and $q$,
and thus,
one can compute the factorization of $x$
by solving SAT on $\F{1}$.

Here we take the following approach to generate $F_x$:
(i) First design a circuit $C_x$,
which we call a {\em test circuit},
such that
$C_x(\veca,\vecb)$
checks whether $\veca\times\vecb$ $=$ $x$.
(ii) Then convert it into a conjunctive form formula $F_x$.
In fact,
there is a standard way to transform a circuit to a conjunctive form formula
(see Lemma~\ref{l:c-to-f}),
by which we can construct a conjunctive form formula $F_x$
with the following property:

\[
\begin{array}{l}
[~C_x(\veca,\vecb)=1~]~~
(\iff~[~(\veca)\times(\vecb)=x~])\cr
\iff~
\exists u_1,...,u_t~
[~F_x(\veca,\vecb,u_1,...,u_t)=\mbox{true}~].
\end{array}
\]

\noindent
Clearly,
this $F_x$ is also good enough for our purpose.
Furthermore,
the size of $F_x$,
i.e.,
the number of variables and clauses,
are closely related to the number of gates of the circuit $C_x$.
Thus,
our goal is now to design a test circuit $C_x$ with small number of gates.

We can easily think of $O(\ell^2)$ size circuit
that multiplies two $\ell$ bit numbers,
which gives a test circuit $\Ctrivx$ of almost the same size.
For the multiplication,
asymptotically the best one (so far) is obtained
by using the Fast Fourier Transform
(\cite{schonhage-strassen:Comput71};
see also \cite{knuth:book81}).
By using this algorithm,
we can design $\Cfft$ with $O(\ell\log\ell\log\log\ell)$ gates.
Unfortunately,
though,
due to its large constant factor,
the size of circuits
(and thus formulas)
obtained in this way become quite large in practice.

In this paper,
we construct test circuits based on the Chinese Remainder Theorem.
Let $m_1,...,m_k$ be relatively prime numbers,
and let $m=m_1\cdot m_2\cdots m_k$.
The Chinese Remainder Theorem claims that
for any $x_1,...,x_k$ such that $0\le x_i<m_i$ for each $i$,
there exists unique $y$, $0\le x<m$, such that
$x\mod m_i$ $=$ $x_i$ for all $i$, $1\le i\le k$.
The following fact is immediate from this claim.

\begin{fact}\label{f:CRT}
For any $x\ge0$ of $2\ell$ bit number,
let $m_1,...,m_k$ be relatively prime numbers
such that $m=m_1\cdot m_2\cdots m_k$ $\ge$ $2^{2\ell}$.
For any $p,q$ of $\ell$ bit number,
$p\times q$ $=$ $x$
if and only if
$p\times q\equiv x$ $\pmod{m_i}$ for all $i$, $1\le i\le k$.
\end{fact}

Let $m_1,...,m_k$ be relatively prime numbers
such that $m=m_1\cdot m_2\cdots m_k$ $\ge$ $2^{2\ell}$ for our $x$.
(Recall $x$ is the product of two $\ell$ bit prime numbers.)
Then we may consider the following circuit $\C{2}$
that checks whether $u\times v=x$,
for given two numbers $u=(\veca)$ and $v=(\vecb)$.

\medskip\noindent\hangindent20pt
(Step 1)~
For every $i$, $1\le i\le k$,
compute $u_i=u\mod m_i$ and $v_i=v\mod m_i$.
(Also for every $i$, $1\le i\le k$,
let $x_i=x\mod m_i$.
Note that
these $x_i$'s are constants and we do not have to compute them.)

\noindent\hangindent20pt
(Step 2)~
For every $i$, $1\le i\le k$,
check whether $u_i\times v_i$ $\equiv$ $x_i$ $\pmod{m_i}$.
If all of them hold,
then output 1;
otherwise,
output 0.

\medskip\noindent
Since the length of each $u_i$ and $v_i$ is much smaller than
that of $u$ and $v$,
we may expect to reduce the complexity of checking.
Note,
however,
it is now necessary to compute each $u_i$ and $v_i$,
which is not so cheap in general.
Also we need to compute $u_i\cdot v_i$ modulo $m_i$.

Here we use integers of the form $2^{e_i}-1$ for each $m_i$.
Then we can reduce
the cost of computing $u_i$, $v_i$, and $u_i\cdot v_i\mod m_i$.
As explained below (Claim~\ref{c:mod}),
we can compute each $u_i$ (resp., $v_i$)
by some $O(\ell)$-size circuit.
Also it will be shown later (Claim~\ref{c:mult-both})
that the cost of computing $u_i\cdot v_i\mod m_i$
is almost the same as that of ordinary multiplication;
hence,
this task can be done by $O(e_i^2)$-size circuit
because both $u_i$ and $v_i$ are $e_i$ bit integers.

Note also that
the relative primality of $2^e-1$ and $2^{e'}-1$
is coincide with $e$ and $e'$
(see Fact~\ref{f:relativeprime} below).
Thus,
we can use $e_1=\lceil\ell/2\rceil$ and $e_2=\lceil\ell/2\rceil+1$.
On the other hand,
if we want to divide the checking into small pieces,
we may choose
the first $k$ prime numbers for $e_1$, $e_2$, ..., $e_k$
such that
$e_1+e_2+\cdots+e_k$ $>$ $\ell+k$
(where $+k$ is for some margin).
In this case,
we can bound $k$ and $e_k$
by $O(e_k/\log e_k)$ and $O((\ell\log\ell)^{1/2})$ respectively,
and thus,
the size of the test circuit $\C{2}$ is
bounded by $O(\ell^{3/2}(\log\ell)^{1/2})$
\cite{horie:al97}.

\begin{fact}\label{f:relativeprime}
For any $e,e'\ge1$,
$2^e-1$ and $2^{e'}-1$ are relatively prime
if and only if
so are $e$ and $e'$.
\end{fact}

Now to get an asymptotically better bound,
we consider applying the Chinese Remainder Theorem recursively.
That is,
we break down
the test of $u_i\times v_i\equiv x_i$ $\pmod{m_i}$ yet further.
Unfortunately,
however,
the characterization like Fact~\ref{f:CRT} does not hold in general.
For example,
while we have
$12\times12$ $\equiv$ $20$ $\pmod{2^5-1}$,
$12$ $\equiv$ $5$ $\pmod{2^3-1}$, and
$20$ $\equiv$ $6$ $\pmod{2^3-1}$,
it does not hold that
$5\times5$ $\equiv$ $6$ $\pmod{2^3-1}$.
Here we extend Fact~\ref{f:CRT} as follows.

\begin{fact}\label{f:newCRT}
For any $n\ge1$ of $e$ bit number,
let $m_1,...,m_k$ be relatively prime numbers
such that $m=m_1\cdot m_2\cdots m_k$ $\ge$ $2^{2e}$.
Then for any $u$, $v$, and $y$, $0\le u,v,y<n$,
we have
$u\times v$ $\equiv$ $y$ $\pmod{n}$
if and only if

\[
\begin{array}{l}
\exists w:\,0\le w<2^{2e}\cr
\displaystyle
\left[~
w\equiv y~\pmod{n}{\rm~~and~~}
\bigwedge_{1\le i\le k}u\times v\equiv w~\pmod{m_i}.
~\right]
\end{array}
\]
\end{fact}

For any number $y$,
and for any $e$ such that $y<2^e$,
we define a circuit $\Crec_{y,e}$ that checks
whether $u\times v$ $\equiv$ $y$ $\pmod{2^e-1}$.
(We will see that
$\Crec_{x,2\ell}$ can be used as a test circuit.)
Intuitively,
for given $u$ and $v$,
we may consider that
$\Crec_{y,e}$ achieves the following {\em nondeterministic} computation.

\medskip\noindent
Let $e_1,...,e_k$ be relatively prime numbers such that
$(2^{e_1}-1)\cdots(2^{e_k}-1)$ $\ge$ $2^{2e}$.

\noindent\hangindent20pt
(Step 1)~
Guess $w$, $0\le w<2^{2e}$,
and check whether $w\equiv y$ $\pmod{2^e-1}$.

\noindent\hangindent20pt
(Step 2)~
For every $i$, $1\le i\le k$,
compute $u_i=u\mod(2^{e_i}-1)$,
$v_i=v\mod(2^{e_i}-1)$, and $w_i=w\mod(2^{e_i}-1)$.

\noindent\hangindent20pt
(Step 3)~
For every $i$, $1\le i\le k$,
check whether $u_i\times v_i$ $\equiv$ $w_i$ $\pmod{2^{e_i}-1}$
by using $\Crec_{w_i,e_i}$.
If all of them hold,
then output 1;
otherwise,
output 0.

\medskip
We consider that
$\Crec_{y,e}$ {\em accepts} $u$ and $v$
if it outputs 1 on some guess $w$.
Formally,
$\Crec_{y,e}$ is a circuit
with some additional input gates for $w$,
and $\Crec_{y,e}$ accepts $u$ and $v$
if and only if
$\Crec_{y,e}(u,v,w,w')$ $=$ $1$ for some $w$ and $w'$.
(Input $w'$ is used
for nondeterministic guesses in the recursive computation.)
Then,
it follows from Fact~\ref{f:newCRT} that
$u\times v$ $\equiv$ $y$ $\pmod{2^e-1}$ holds
if and only if
$\Crec_{y,e}(u,v,w,w')$ $=$ $1$ for some $w$ and $w'$.

In order to determine $\Crec_{y,e}$ precisely,
we need to define $k$ and the way to select $e_1,...,e_k$.
Here we define $k=k(e)$
by using some unbounded but slowly increasing function $k$,
e.g., $k(e)=\log e$.
For $e_1<...<e_k$,
we choose the smallest $k$ primes larger than $(2e+k)/k$.
Then we have
$(2^{e_1}-1)\cdots(2^{e_k}-1)$ $\ge$ $2^{2e}$.
It is easy to see that
our choice of parameters yields a circuit achieving the desired test.

\begin{lemm}\label{l:rec}
The size of $\Crec_{y,e}$ is $O(e^{1+\varepsilon})$ for any $\varepsilon>0$.
\end{lemm}

\beginproof
Here we fix any $\varepsilon>0$,
and show that there exists some constant $c$ such that
$\csize(\Crec_{y,e})$ $\le$ $c\cdot e^{1+\varepsilon}$
for sufficiently large $y$ and $e$.
In the following discussion,
let us also fix $y$ and $e$.

First we give an upper bound for computing $u\mod(2^f-1)$ for a given $u$.
Although results are from $0$ to $2^f-2$,
we allow to use $2^f-1$,
which is regarded as $0$.
Thus,
the binary representation of $0$ is
either $(0,0,...,0)$ or $(1,1,...,1)$.
We call
this slightly relaxed way to represent numbers modulo $2^f-1$
as an {\em extended binary representation}.
The notation $u\emod(2^f-1)$ is used to denote
$u\mod(2^f-1)$ representing the extended binary representation.
In order to distinguish from $(1,1,...,1)$,
we call $(0,0,...,0)$ as the {\em real 0 representation}.

For our analysis,
we need the following claims.
(The claim proved as a special case of
the corresponding one in Section~3.
Thus,
we omit its proof.)

\begin{claim}\label{c:mod}
For any $f\ge1$,
we can construct a circuit $\cmod{e,f}$ with the following properties.

\itemplain{(1)}
$\cmod{e,f}$ is an $e$ input and $f$ output circuit.

\itemplain{(2)}
On input $u$, $0\le u\le 2^e-1$,
$\cmod{e,f}(u)$ yields $u\emod(2^f-1)$.
Also the output becomes the real 0 representation
if and only if $u=0$.

\itemplain{(3)}
The size of $\cmod{e,f}(u)$ is
bounded by $c_1\cdot e$ for some constant $c_1$.
\end{claim}

Now we show,
by induction on $e$,
that $\csize(\Crec_{y,e})$ $\le$ $c\cdot e^{1+\varepsilon}$.
From the outline of $\Crec_{y,e}$,
we have the following bound.

\[
\begin{array}{lcl}
\csize(\Crec_{y,e})
&=&
\displaystyle
\sum_{i=1}^k
\left(\csize(\Crec_{y_i,e_i})
+2\csize(\cmod{e,e_i})+\csize(\cmod{2e,e_i})\right)+\csize(\cmod{2e,e})+k\cr
&\le&
\displaystyle
\sum_{i=1}^k
\left(c\cdot e_i^{1+\varepsilon}+2c_1\cdot e+c_1\cdot2e\right)+c_1\cdot2e+k
~\le~
\sum_{i=1}^k
c\cdot e_i^{1+\varepsilon}+c_2\cdot ke.
\end{array}
\]

\noindent
Here the term $+k$ is for the number of AND gates
that summarize the check at (Step1) and (Step3).

Recall that we assume that
$k$ is determined by a slowly growing function,
and that $e_1<e_2<\cdots<e_k$ are
the smallest $k$ primes larger than $(2e+k)/k$.
Hence,
by using the Prime Number Theorem,
we can bound $e_k$ by $3e/k$ (for sufficiently large $e$).
Thus,
we have

\[
\csize(\Crec_{y,e})
~\le~ck\cdot e_k^{1+\varepsilon}+c_2\cdot ke
~\le~ck\left({3e\over k}\right)^{1+\varepsilon}+c_2\cdot e^{1+\varepsilon},
\]

\noindent
which is bounded by $ce^{1+\varepsilon}$
if $k$ (i.e., $k(e)$) is large enough.
\endproof

Finally,
we define a SAT instance $\Frecx$.
Precisely speaking,
$\Crec_{x,2\ell}$ is not a test circuit $C_x$;
but $C_x(u,v)=1$ if and only if
the partially assigned circuit $\Crec_{x,2\ell}(u,v,-,-)$ is satisfiable.
Hence,
the standard transformation
from circuits to conjunctive normal form formulas (Lemma~\ref{l:c-to-f})
yields a SAT instance $\Frecx$ with the desired property.
Furthermore,
the size of $\Frecx$ is almost the same as that of $\Crec_{x,2\ell}$.
Therefore,
the following theorem holds.

\begin{theo}\label{t:rec}
For any $\varepsilon>0$,
we can construct SAT instances
with $O(\ell^{1+\varepsilon})$ variables and clauses
(in the conjunctive normal form)
that are as hard as factorizing the product of two $\ell$ bit prime numbers.
\end{theo}


\smallsection
Concrete Examples

Here we examine
the applicability of our method with some concrete examples,
i.e.,
the cases where $\ell=30$, $40$, ... .
For such examples,
to reduce the size of formulas,
we need some small techniques different from the previous section;
in fact,
the recursive application of the Chinese Remainder Theorem does not work
due to its large constant factor.

First we state our construction,
and then estimate the size of obtained Boolean formulas.
Here we follow the same approach as Section~2;
that is,
for any $x$,
a product of two $\ell$ bit prime numbers $p$ and $q$,
we first define a test circuit and transform it to a SAT instance.
We fix $x$, $p$, and $q$ in the following discussion.

The key task is to test whether $u\times v=x$ for given $u$ and $v$.
By using the Chinese Remainder Theorem,
we divide this test into small pieces of similar tests.
Since we cannot apply the Chinese Remainder Theorem recursively,
we would like to divide the test as small pieces as possible.
For example,
we may choose the smallest $k$ prime numbers $e_1,...,e_k$
such that $e_1+\cdots+e_k$ $\ge$ $2\ell+k$ and achieve the test
by checking whether $u_i\times v_i\equiv x_i$ $\pmod{m_i}$
for all $i$, $1\le i\le k$,
where $m_i=2^{e_i}-1$, $u_i=u\mod m_i$, $v_i=v\mod m_i$, and $x_i=x\mod m_i$.
Our main idea here is
to use $m'_i=2^{e_i}+1$ as well as $m_i=2^{e_i}-1$.
We also use $m_0=2^{e_0}$ for some $e_0\ge1$.
(In the following,
we let $u'_i=u\mod m'_i$, $v'_i=v\mod m'_i$, $x'_i=x\mod m'_i$,
$u_0=u\mod m_0$, $v_0=v\mod m_0$, and $x_0=x\mod m_0$.)

Note that
for any $e$,
one of $2^e-1$ and $2^e+1$ is divisible by 3;
but 3 is the largest common factor of $2^{e}\pm1$ and $2^{e'}\pm1$
for any $e$ and $e'$, $e\ne e'$.
Also $2^e$ is relatively prime with any $2^{e'}\pm1$.

\begin{fact}
For any relatively prime numbers $e,e'\ge2$,
$\gcd(2^e\pm1,2^{e'}\pm1)$ $=$ 1 or 3.
(Clearly,
$\gcd(2^e-1,2^e+1)$ $=$ 1.)
\end{fact}

We note that
the Chinese Remainder Theorem (i.e., Fact~\ref{f:CRT}) works
if $\gcd(m_0,m_1,...,m_k,m'_1,...,m'_k)\ge2^{\ell}$.
Hence,
roughly speaking,
it is enough to choose relatively prime numbers $e_1,...,e_k$ and some $e_0$
such that
$2(e_1+\cdots+e_k)+e_0-k\log3$ $>$ $2\ell$.
Clearly,
this idea enables us to choose smaller modulos.
Furthermore,
there is another advantage of
using both $m_i=2^{e_i}-1$ and $m'_i=2^{e_i}+1$.
As we see below (Claim~\ref{c:mod-both}),
the most of
the computation of $u_i=u\mod m_i$ and $u'_i=u\mod m'_i$ can be shared,
and $u'_i$ is computable almost as a byproduct of $u_i$.
It is also shown (Claim~\ref{c:mult-both})
that the multiplication cost modulo $m'_i$
is almost the same as the multiplication cost modulo $m_i$.

To summarize,
we choose $e_0,e_1,...,e_k$ so that
$\gcd(m_0,m_1,...,m_k,m'_1,...,m'_k)\ge2^{\ell}$,
and construct $\Ccexx$ that tests
whether $u\times v=x$ for given inputs $u$ and $v$ in the following way.

\medskip\noindent\hangindent20pt
(Step 1)~
Compute $u_i,u'_i,v_i$, and $v'_i$ for every $i$, $1\le i\le k$.
(Note that
$u_0$ (resp., $v_0$) is just the last $e_0$ bits of $u$ (resp., $v$),
and hence,
we do not need to compute them.)

\noindent\hangindent20pt
(Step 2)~
Check whether $u_i\times v_i$ $\equiv$ $x_i$ $\pmod{m_i}$
and $u'_i\times v'_i$ $\equiv$ $x'_i$ $\pmod{m'_i}$
for every $i$, $1\le i\le k$,
and also check whether $u_0\times v_0$ $\equiv$ $x_0$ $\pmod{m_0}$.
If all of them hold,
then output 1;
otherwise,
output 0.

\medskip
Now we estimate the size of $\Ccexx$ in detail.
First we remark on the type of gates used in circuits.
Though it is standard to construct circuits by using 2-fan-in gates,
here we also use 3-fan-in gates,
since 3-fan-in gates are useful for addition and subtraction.
Clearly,
we can reduce circuit size
by using $k$-fan-in gates for larger $k$;
but the number of clauses in the conjunctive form
grows proportionally in $2^k$.
Here by using 3-fan-in gates,
we can not only simplify our argument,
but also we can reduce the total number of clauses in the conjunctive form.
In the following,
in order to distinguish
the number of 2-fan-in and 3-fan-in gates,
we write,
e.g., $\csize(C)$ $=$ $\fbox{320}+1500$,
by which we mean that
$C$ consists of 320 3-fan-in gates and 1500 2-fan-in gates.

First we state a precise relationship between
a circuit $C$ and
a SAT instance $F$ transformed from $C$ by the standard reduction.

\begin{lemm}\label{l:c-to-f}
Let $C$ be a circuit
with $n$ inputs, $s_1$ fan-in-2 gates, and $s_2$ fan-in-3 gates;
let $m=s_1+s_2$.
From this $C$,
we can construct a formula $F$ in the extended conjunctive form
with $n+m$ variables and $m$ clauses
that simulates $C$ in the following sense:

\[
\begin{array}{l}
[~C(a_1,...,a_n)=1~]~
\iff~
\exists u_1,...,u_m~
[~F(a_1,...,a_n,u_1,...,u_m)=\mbox{true}~].
\end{array}
\]

\noindent
The formula can be transformed into the 4-conjunctive normal form
with at most $4s_1+8s_2$ clauses.
\end{lemm}

Next we prepare circuits for some basic arithmetic operations.

\begin{claim}\label{c:inc}
The addition of one bit number to $e$ bit number is computable
by a circuit $\cinc{e}$ with $\csize(\cinc{e})=2e$.
We use $\sizeinc{e}$ to denote this circuit size.
\end{claim}

\beginproof
The circuit $\cinc{e}$ is defined as Figure~1 below.
Here gates with label $\oplus$ are exclusive-or gates.
\endproof

\begin{center}
\psbox[width=0.8\textwidth]{fig1.ps}~\\
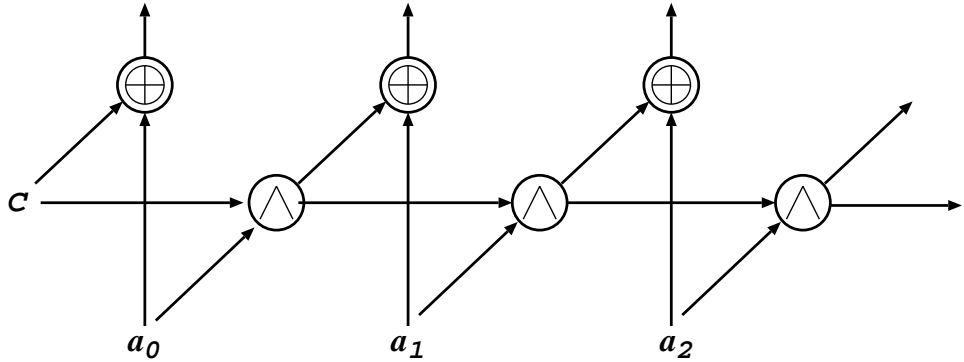
{\bf Fig. 1:}~
Circuit $\cinc{e}$
\end{center}

\begin{claim}\label{c:add}
The addition of two $e$ bit numbers is computable
by a circuit $\cadd{e}$ with $\csize(\cadd{e})=\myfbox{2e}$.
We use $\sizeadd{e}$ to denote this circuit size.
\end{claim}

\beginproof
The circuit $\cadd{e}$ is defined as Figure~2 below.
Here gates with label C are gates computing
the current bit from two input bits and a carry.
\endproof

\begin{center}
\psbox[width=0.8\textwidth]{fig2.ps}~\\
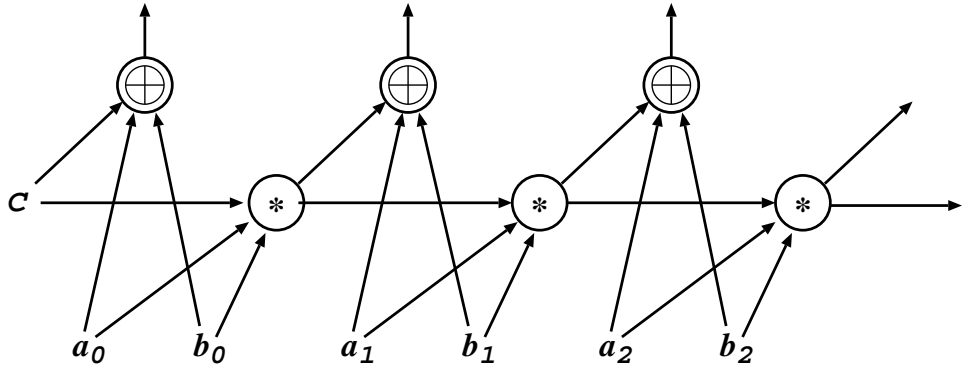
{\bf Fig. 2:}~
Circuit $\cadd{e}$
\end{center}

\begin{claim}\label{c:sub}
The subtraction of two $e$ bit numbers is computable
by a circuit $\csub{e}$ with $\csize(\csub{e})=\myfbox{2e}$.
More precisely,
$\csub{e}$ takes two $e$ bit numbers $u$ and $v$ as input,
and outputs $(u-v)\mod{2^e}$ and
$c$ indicating whether $u-v\ge0$
($c=0$ if $u-v\ge0$, and $c=1$ if otherwise).
We use $\sizesub{e}$ to denote this circuit size.
\end{claim}

\begin{claim}\label{c:mod-both}
We can construct a circuit $\cmod{e}$ with the following properties.

\itemplain{(1)}
$\cmod{e}$ is an $\ell$ input and $2e+1$ output circuit.

\itemplain{(2)}
On input $u$,
$\cmod{e}(u)$ yields $u\emod(2^e-1)$ and $u\mod(2^e+1)$
at the first $e$ output gates and the last $e+1$ gates respectively.

\itemplain{(3)}
The size of $\cmod{e}$ is $\myfbox{2\ell+2e}$ $+$ $4e+2\ell'$,
where $\ell'=\ell-(\ell\mod e)$.
\end{claim}

\beginproof
Let $u$ be $\ell$ bit number,
for which we want compute $s=u\emod(2^e-1)$ and $t=u\mod(2^e+1)$.
Let $(u_0,...,u_{h-1})$ be its base $2^e$ representation.
That is,
$u$ $=$ $u_0+u_12^e+u_22^{2e}+\cdots+u_{h-1}2^{(h-1)e}$,
where $h=\lceil\ell/e\rceil$.
Here we assume that
$h-1$ is even and $h-1=2h'$ for some $h$.
(The odd case is treated similarly.)
Then we have

\[
\begin{array}{lcl}
s&=&(u_0+u_1+u_2+u_3\cdots+u_{2h'})\emod(2^e-1)\cr
 &=&((u_0+u_2+\cdots+u_{2h'})+(u_1+u_3+\cdots+u_{2h'-1}))\emod(2^e-1),
{\rm~~and}\cr
t&=&(u_0-u_1+u_2-u_3+\cdots+u_{2h'})\mod(2^e+1)\cr
 &=&((u_0+u_2+\cdots+u_{2h'})-(u_1+u_3+\cdots+u_{2h'-1}))\emod(2^e+1).
\end{array}
\]

Note also that for any $x,y$, $0\le x,y\le 2^e$,
we have

\[
\begin{array}{lcl}
(x+y)\emod(2^e-1)&=&(x+y)\mod2^e+c_{x,y},{\rm~~and}\cr
(x+y)\emod(2^e+1)&=&(x+y)\mod2^e-c_{x,y},
\end{array}
\]

\noindent
where $c_{x,y}$ is the $(e+1)$th bit of $x+y$,
or the $e$th carry of $x+y$.

These observations suggests us
to compute the following $v_+$ and $v_-$.

\[
\begin{array}{lcl}
v_+&=&
((\cdots((u_0+u_2)\mod2^e+u_4+c_3)\mod2^e+\cdots)
+u_{2h'}+c_{2h'-1})\mod2^e,{\rm~~and}\cr
v_-&=&
((\cdots((u_1+u_3+c_2)\mod2^e+u_5+c_4)\mod2^e+\cdots)
+u_{2h'}+c_{2h'-2})\mod2^e,
\end{array}
\]

\noindent
where $c_i$ is the $e$th carry of
the addition of a partial sum $\alpha_{i-2}$ and $u_i+c_{i-1}$.
The following figure illustrates this computation.

\begin{center}
\begin{tabular}{rl}
         \fbox{\hbox to 13mm{\hfil$u_0$\hfil}}\\
   $+$   \fbox{\hbox to 13mm{\hfil$u_2$\hfil}}\\[2pt]\hline
 $c_2$   \fbox{\hbox to 13mm{\hfil$\alpha_2$\hfil}}\\
   $+$   \fbox{\hbox to 13mm{\hfil$u_4$\hfil}}&$\leftarrow c_3$\\[2pt]\hline
\multicolumn{1}{c}{~~~~~~$\vdots$}\\\hline
 $c_{2h'-2}$
         \fbox{\hbox to 13mm{\hfil$\alpha_{2h'-2}$\hfil}}\\
   $+$   \fbox{\hbox to 8mm{\hfil$u_{2h'}$\hfil}}&$\leftarrow c_{2h'-1}$\\[2pt]\hline
$c_{2h'}$\fbox{\hbox to 13mm{\hfil$v_+$\hfil}}
\end{tabular}
~~~~~~~
\begin{tabular}{rl}
         \fbox{\hbox to 13mm{\hfil$u_1$\hfil}}\\
   $+$   \fbox{\hbox to 13mm{\hfil$u_3$\hfil}}&$\leftarrow c_2$\\[2pt]\hline
 $c_3$   \fbox{\hbox to 13mm{\hfil$\alpha_3$\hfil}}\\
   $+$   \fbox{\hbox to 13mm{\hfil$u_5$\hfil}}&$\leftarrow c_4$\\[2pt]\hline
\multicolumn{1}{c}{~~~~~$\vdots$}\\\hline
 $c_{2h'-3}$
         \fbox{\hbox to 13mm{\hfil$\alpha_{2h'-3}$\hfil}}\\
   $+$   \fbox{\hbox to 13mm{\hfil$u_{2h'-1}$\hfil}}&$\leftarrow c_{2h'-2}$\\[2pt]\hline
$c_{2h'-1}$
         \fbox{\hbox to 13mm{\hfil$v_-$\hfil}}
\end{tabular}\\[5mm]
{\bf Fig. 3:}~
Computation of $v_+$ and $v_-$.
\end{center}

Then it is easy to see that $s$ and $t$ are obtained by

\[
\begin{array}{lcl}
s&=&(s_++s_-+c_{2h'})\mod2^e+c_+,{\rm~~and}\cr
t&=&(s_+-s_--c_{2h'})\mod2^e+c_-,
\end{array}
\]

\noindent
where $c_+$ and $c_-$ are respectively
the $e$th carry of $s_++s_-+c_{2h'}$ and
the negative $e$th carry of $s_+-s_--c_{2h'}$.

Our circuit $\cmod{e}$ is defined following this outline.
Recall that
$\cadd{e}$ can be modified with no additional gate
for adding two numbers with a carry
(Claim~\ref{c:add});
the same property holds for $\csub{e}$.
Thus,
the size of $\cmod{e}$ is estimated as follows.

\[
\begin{array}{lcl}
\csize(\cmod{e})
&=&(2h'-1)\sizeadd{e}+\sizeadd{\ell''}+\sizeadd{e}+\sizesub{e}
   +\sizeinc{\ell'}+2\sizeinc{e}\cr
&=&(h-2)\myfbox{2e}+\myfbox{2\ell''}+\myfbox{4e}+2\ell'+4e
~=~\myfbox{2\ell+2e}+4e+2\ell'.
\end{array}
\]

\noindent
Here $\ell''=\ell\mod e$ and $\ell'=\ell-\ell''$.
Note that
adding $u_{2h'}$ to the partial sum is computed
with two circuits $\cadd{\ell''}$ and $\cinc{\ell'}$.
\endproof

\begin{claim}\label{c:mult-both}
For any $e\ge1$,
we can construct a circuit $\cmult{e}$ and $\cmultd{e}$
with the following properties.

\itemplain{(1)}
$\cmult{e}$ is $2e$ input and $e$ output circuit,
and $\cmultd{e}$ is $2(e+1)$ input and $e+1$ output circuit.

\itemplain{(2)}
For any pair of input integers $u$ and $v$, $0\le u,v\le 2^e-1$,
$\cmult{e}$ computes\break
$u\cdot v\emod(2^e-1)$.
Similarly,
for any pair of input integers $u$ and $v$, $0\le u,v\le 2^e+1$,
$\cmultd{e}$ computes $u\cdot u\mod(2^e+1)$.

\itemplain{(3)}
The size of $\cmult{e}$ and $\cmultd{e}$ are bounded by
$\myfbox{2(e-1)e}+e^2+2e$ and $\myfbox{2e^2+e+1}+e^2+4e$ respectively.
\end{claim}

\noindent
\begin{minipage}{0.55\textwidth}
\beginproof
First we consider $\cmult{e}$.
Consider any integers $u,v$, $0\le u,v\le 2^e-1$;
let $(a_{e-1},...,a_0)$ and $(b_{e-1},...,b_0)$ be
binary representations of $u$ and $v$ respectively.
Intuitively,
$w$ $=$ $u\cdot v\mod(2^e-1)$ is computed as in Figure~4.
More specifically,
it is computed as (1) below.
Here $u_i$ $=$ $(a_{e-i-1},...,a_0,a_1,...,a_{e-i})\times b_i$;
that is,
each bit of $u_i$ is computed as $a_j\land b_i$.
Hence,
for computing $w$,
we need $e-1$ $\cadd{e}$ circuits, one $\cinc{e}$ circuit,
and $e^2$ AND gates.
\end{minipage}
~~~~~
\begin{minipage}{0.35\textwidth}
\begin{center}
\begin{tabular}{rl}
         $a_{e-1}a_{e-2}\cdots a_1a_0$\\[-1pt]
$\times$ $b_{e-1}b_{e-2}\cdots b_1b_0$\\\hline
         $a_{e-1}a_{e-2}\cdots a_1a_0$&$\times b_0$\\[-1pt]
         $a_{e-2}\cdots a_1a_0a_{e-1}$&$\times b_1$\\[-1pt]
\multicolumn{1}{c}{$\vdots$}\\[-1pt]
$+$      $a_0a_{e-1}a_{e-2}\cdots a_1$&$\times b_1$\\\hline
         \fbox{\hbox to 26mm{\hfil$w$\hfil}}
\end{tabular}\\[1mm]
* Carries are omitted here.\\[3mm]
{\bf Fig. 4:}~
$u\cdot v\mod(2^e-1)$
\end{center}
\end{minipage}

\begin{equation}
w~=~
((\cdots(((u_0+u_1)\mod2^e+u_2+c_1)\mod2^e)\cdots)+u_{e-1}+c_{e-2})\mod2^e,
\end{equation}

Thus,
the size of $\cmult{e}$ is estimated as follows.
\[
\csize(\cmult{e})
~=~(e-1)\sizeadd{e}+\sizeinc{e}+e^2
~=~\myfbox{2(e-1)e}+e^2+2e.
\]

Next define circuit $\cmultd{e}$.
This time $u$ and/or $v$ can be $2^e$.
Hence,
we need to represent them
as $(a_e,...,a_0)$ and $(b_e,...,b_0)$;
but let us also consider $u'=(a_{e-1},...,a_0)$ and $v'=(b_{e-1},...,b_0)$.
Then we have

\[
w
~=~u\cdot v\mod(2^e+1)
~=~(u'\cdot v'\mod(2^e+1)-(u''+v'')+a_e\cdot b_e)\mod(2^e+1),
\]

\noindent
where $u''=u'\cdot b_e$ and $v''=v'\cdot a_e$.

We first consider how to compute $u'\cdot v'\mod(2^e+1)$.
Just compute $u'\cdot v'$ in the standard way,
which gives us $2e$ bit number.
Let $w_-$ and $w_+$ denote numbers
at the first $e$ bits and the last $e$ bits respectively.
Then we have $u'\cdot v'\mod(2^e+1)$ $=$ $(w_+-w_-)\mod2^e+c$,
where $c$ is the negative $e$th carry of $w_+-w_-$.
Thus,
$w$ is obtained by
$(w_+-(w_-+u''+v''))\mod2^e+c+a_e\cdot b_e$.
Notice here that
at most one of $w_-$, $u''$, $v''$ is nonzero.
Hence,
$w_-+u''+v''$ is computable by bit-wise or,
which can be done by $e$ 3-fan-in OR gates.
Similarly,
if $a_e\cdot b_e=1$,
then the other term for $w$ is zero.
Thus,
the size of our circuit $\cmultd{e}$,
which computes $w$ following this outline,
is estimated as follows.

\[
\begin{array}{lcl}
\csize(\cmultd{e})
&=&\mbox{(\# of gates for $u'\cdot v'$)}
   +\mbox{(\# of gates for $u''$ and $v''$)}\cr
& &+\mbox{(\# of gates for $w_-+u''+v''$)}
   +\sizesub{e}+\sizeinc{e}\cr
& &+\mbox{(\# of gates for $+a_e\cdot b_e$)}\cr
&=&\myfbox{2(e-1)e}+e^2+2e+\myfbox{e}+\myfbox{2e}+2e+\myfbox{1}\cr
&=&\myfbox{2e^2+e+1}+e^2+4e.
\end{array}
\]
\square\paragraphskip

Now the size of our test circuit $\Ccexx$,
which uses these circuits,
is estimated as follows.

\begin{lemm}\label{l:concrete}
The circuit $\Ccexx$ outlined above tests
whether $u\cdot v=x$ for given inputs $u$ and $v$,
and we can bound its size as follows,
where $\ell$ is the length of $x$'s prime factors,
$e_0,...,e_k$ are parameters defined above,
and $\ell'_i=\ell-\ell\mod e_i$, $1\le i\le k$.

\[
\begin{array}{lcl}
\csize(\Ccexx)
&\le&\myfbox{\displaystyle
             \sum_{i=1}^k(4e_i^2+3e_i)+e_0^2-e_0+4k\ell+k}\cr
&   &\displaystyle
     +\sum_{i=1}^k(2e_i^2+16e_i+2\ell'_i)+e_0^2/2+e_0/2-2.
\end{array}
\]
\end{lemm}

\beginproof
It follows from the above outline
that $\Ccexx$ consists of,
(i) for each $i$, $1\le i\le k$,
two $\cmod{e_i}$, one $\cmult{e_i}$, and one $\cmultd{e_i}$ circuits,
(ii) a circuit for computing $u_0\times v_0\mod2^{e_0}$, and
(iii) gates for checking every obtained product is equal to $x_i$.
It is not easy to see that
a circuit for $u_0\times v_0\mod2^{e_0}$
requires $\myfbox{(e_0-1)e_0}+(e_0-1)e_0/2$ gates,
and that the whole equality check can be done
with $\displaystyle e_0-1+\sum_{i=1}^k(2e_i-1)+k-1$ gates.
Hence,
we have

\[
\begin{array}{lcl}
\csize(\Ccexx)
&=&
\displaystyle
\sum_{i=1}^k
\left(2\csize(\cmod{e_i})+\csize(\cmult{e_i})+\csize(\cmultd{e_i})\right)\cr
& &
+\myfbox{(e_0-1)e_0}+(e_0-1)e_0/2+e_0-1+\sum_{i=1}^k(2e_i-1)+k-1\cr
&=&
\displaystyle
\sum_{i=1}^k
\left(\myfbox{4\ell+4e_i+2(e_i-1)e_i+2e_i^2+e_i+1}
      +8e_i+4\ell'_i+e_i^2+2e_i+e_i^2+4e_i\right)\cr
& &
\displaystyle
+\myfbox{(e_0-1)e_0}+(e_0-1)e_0/2+e_0-1+\sum_{i=1}^k(2e_i-1)+k-1\cr
&=&
\displaystyle
\myfbox{\displaystyle
\sum_{i=1}^k(4e_i^2+3e_i)+e_0^2-e_0+4k\ell+k}
+\sum_{i=1}^k(2e_i^2+16e_i+2\ell'_i)+e_0^2/2+e_0/2-2.
\end{array}
\]
\square\paragraphskip

\begin{theo}\label{t:concrete}
For a given $x$,
a product of two $\ell$ bit prime numbers,
we can construct a SAT instance $\Fcexx$ that is as hard as factorizing $x$,
and that has at most the following number of variables,
where $e_0,...,e_k$ and $\ell'_1,...,\ell'_k$ are parameters defined above.

\[
\sum_{i=1}^k(6e_i^2+19e_i+2\ell'_i)+3e_0^2/2-e_0/2+4k\ell+k+2\ell-2.
\]

\noindent
$\Fcexx$ has at most this number of clauses in the extended 4-conjunctive form
and at most
$\sum_{i=1}^k(40e_i^2+88e_i+8\ell'_i)+10e_0^2-6e_0+32k\ell+8k-8$ clauses
in the 4-conjunctive normal form.
\end{theo}

Now we estimate the size of formulas for several concrete cases.
For comparison,
let us also estimate
the size of the formula $\Ftrivx$ obtained from $x$
by the straightforward reduction explained in Introduction.
(For our concrete examples,
formulas obtained by using the FFT become much larger than
the ones obtained by the straightforward reduction.)

\begin{prop}\label{p:triv}
For a given $x$,
a product of two $\ell$ bit prime numbers,
the formula $\Ftrivx$ has $3\ell^2+2\ell-1$ variables.
It has about this number of clauses in the extended 4-conjunctive form
and at most $20\ell^2-8\ell-4$ clauses in the 4-conjunctive normal form.
\end{prop}

\beginproof
It is easy to show that
the size of
the straightforward circuit multiplying two $\ell$ bit numbers is
$(\ell-1)\cdot\sizeadd{\ell}+\ell^2$ $=$ $\myfbox{2(\ell-1)\ell}+\ell^2$.
The test circuit needs $2\ell-1$ more gates for checking
whether the obtained product is equal to $x$,
and thus,
its size becomes
$\myfbox{2(\ell-1)\ell}+\ell^2+2\ell-1$.
Then the above bounds follow from Lemma~\ref{l:c-to-f}.
\endproof

Table~1 below shows
the size of $\Fcexx$ and $\Ftrivx$ obtained from $x$,
a product of two $\ell$ bit prime numbers;
that is,
solving SAT problem for $\Fcexx$ and $\Ftrivx$ is as hard as factorizing $x$.
The column ``\# of var.s'' is
for the number of variables of obtained formulas;
hence,
it also bounds
the number of clauses of the formulas in the extended 4-conjunctive form.
On the other hand,
the column ``\# of clauses'' is
for the number of clauses of the formulas in the 4-conjunctive normal form.
For these formulas,
the number of clauses in the 4-conjunctive normal form
is approximately 6 times larger than the number of variables.

\begin{center}
~\\
\begin{tabular}{|r|r|r|r|r|l|}
\hline
&\multicolumn{2}{|c|}{$\Ftrivx$}
&\multicolumn{3}{|c|}{$\Fcexx$}\\
\cline{2-6}
\multicolumn{1}{|c|}{$\ell$}
&\multicolumn{1}{|c|}{\# of var.s}
&\multicolumn{1}{|c|}{\# of clauses}
&\multicolumn{1}{|c|}{\# of var.s}
&\multicolumn{1}{|c|}{\# of clauses}
&\multicolumn{1}{|c|}{$e_0,e_1,...$}\\
\hline
30&2,759&11,756
  &2,767&17,240&16, 4, 5, 7, 9\\
\hline
40&4,879&31,676
  &4,103&25,728&16, 7, 8, 9, 11\\
\hline
50&7,599&49,596
  &5,657&35,776&27, 5, 7, 8, 9, 11\\
\hline
60&10,919&71,516
  &7,315&46,328&23, 5, 7, 8, 9, 11, 13\\
\hline
70&14,839&97,436
  &9,347&59,448&27, 5, 7, 9, 11, 13, 16\\
\hline
128&49,407&326,652
  &22,165&142,344&27, 7, 11, 13, 15, 16,\\
&&&&&             17, 19, 23\\
\hline
256&197,119&1,308,668
  &63,652&406,860&62, 7, 11, 13, 17, 19,\\
&&&&&             23, 25, 27, 29, 31, 32\\
\hline
\end{tabular}\\[3mm]
{\bf Table~1:}~
The size of formulas
\end{center}

Consider first
the task of generating test instances for a given SAT algorithm.
From the view point of
the Factorization Problem (FACT),
the case $\ell=30$,
i.e.,
factorizing a product of two 30 bit primes,
is not so difficult.
It is solvable in a few minutes
by a straightforward algorithm on a small workstation.
But the problem suddenly becomes difficult when $\ell>40$.
Thus,
those instances generated with $\ell=40$ or $\ell=50$
would be quite good examples
for testing the performance of SAT algorithms.
Note that
if we use some advanced algorithm like the Quadratic Sieve,
factorization up to $\ell=100$
is computable in one to two hours
on a mid size workstation \cite{kobayashi}.
But it is hard to think of a SAT algorithm
incorporating such a specialized algorithm.

Next analyze the hardness of the SAT
by using our knowledge on the hardness of the FACT.
It has been widely believed (see, e.g., \cite{schneier94})
factorizing 512 bit numbers is hard to solve,
which is the case $\ell=256$.
Now from Table~1,
this corresponds via our reduction
to SAT instances with approximately 63,000 variables.
That is,
some (in fact many) SAT instances with 63,000 variables are intractable.
Notice that
by the straightforward reduction,
we cannot show the same hardness
unless SAT instances have more than 190,000 variables.
In Table 1,
we also estimate the size of SAT instances
generated from 256 bit numbers (i.e., $\ell=128$),
which are still quite difficult to factorize
(i.e., one day task on a mid size workstation \cite{kobayashi})
in practice.


\def\Large{\large}

\end{document}